\documentclass{article}
\usepackage{spconf,amsmath,graphicx,subcaption}
\usepackage[utf8]{inputenc}

\def\psfancypar#1#2{\begingroup\def\par{\endgraf\endgroup\lineskiplimit=0pt}
               \setbox2=\hbox{\large\sc #2}
               \newdimen\tmpht \tmpht \ht2 \advance\tmpht by \baselineskip
               \font\hhuge=cmti12 at \tmpht
              \setbox1=\hbox{{\hhuge #1}}
               \count7=\tmpht \count8=\ht1
               \divide\count8 by 1000 \divide\count7 by \count8
               \tmpht=.001\tmpht\multiply\tmpht by \count7
               \font\hhuge=cmbx10 at \tmpht
               \setbox1=\hbox{{\hhuge #1}}
               \noindent
                \hangindent1.05\wd1
               \hangafter=-2 {\hskip-\hangindent
               \lower1\ht1\hbox{\raise1.0\ht2\copy1}%
                \kern-0\wd1}\copy2\lineskiplimit=-1000pt}

\newcommand{\boxit}[1]{\vbox{\hrule\hbox{\vrule\kern6pt
\vbox{\kern6pt#1 \vspace{-10 pt}\kern6pt}\kern6pt\vrule}\hrule}}
\newcommand{\boxitsl}[1]{\vbox{\hrule\hbox{\vrule\kern6pt
\vbox{\kern6pt#1 \vspace{3 pt}\kern6pt}\kern6pt\vrule}\hrule}}
\def\thetabf{{\mbox{\boldmath$\theta$\unboldmath}}}

\def\reals{ { {\rm  I \kern-0.15em R }  } }
\def\complex{\hbox{ \,{{\rm C} \kern-0.50em \raise0.20ex {  |}}\,}}

\def\Pibf{\hbox{$\boldsymbol{\Pi}$}}

\def\abf{\hbox{\bf a}}

\def\cbf{\hbox{\bf c}}

\def\ebf{\hbox{\bf e}}
\def\fbf{\hbox{\bf f}}

\def\vbf{\hbox{\bf v}}

\def\xbf{\hbox{\bf x}}
\def\ybf{\hbox{\bf y}}

\def\xbf{\hbox{\bf x}}
\def\ybf{\hbox{\bf y}}

\def\Dbf{\hbox{\bf D}}

\def\Ibf{\hbox{\bf I}}

\def\Pbf{\hbox{\bf P}}

\def\Xbf{\hbox{\bf X}}

\def\be{\vskip .3cm \begin{equation}}
\def\ee{\end{equation} \vskip .4cm \noindent}

\newcounter{remarknr}
\renewcommand{\theremarknr}{\arabic{remarknr}}

\newcounter{assumpnr}
\renewcommand{\theassumpnr}{\arabic{assumpnr}}

\newcounter{examplenr} 
\renewcommand{\theexamplenr}{\arabic{examplenr}}

\begin{document}

\title{SEPARABLE DELAY AND DOPPLER ESTIMATION IN PASSIVE RADAR\sthanks{
\tiny The research was funded by the strategic innovation programme ``Smartare Elektroniksystem'', a joint research project financed by VINNOVA, Formas and the Swedish Energy Agency, and by SAAB.}}

\name{Mats Viberg$^{1,2}$, Daniele Gerosa$^{2}$, Tomas McKelvey$^{2}$, Patrik Dammert$^{2, 3}$, Thomas Eriksson$^{2}$}
\address{\small $^1$ Department of Mathematics and Natural Sciences, Blekinge Institute of Technology, Karlskrona, Sweden\\ \small
$^2$ Department of Electrical Engineering, Chalmers University of Technology, Göteborg, Sweden \\ \small
$^3$ SAAB AB, Göteborg, Sweden}

\maketitle

\begin{abstract}
In passive radar, a network of distributed sensors exploit signals from so-called Illuminators-of-Opportunity to detect and localize targets. We consider the case where the IO signal is available at each receiver node through a reference channel, whereas target returns corrupted by interference are collected in a separate surveillance channel. The problem formulation is similar to an active radar that uses a noise-like waveform, or an integrated sensing and communication application. 

The available data is first split into batches of manageable size. In the direct approach, the target's time-delay and Doppler parameters are estimated jointly by incoherently combining the batch-wise data. We propose a new method to estimate the time-delay separately, thus avoiding a costly 2-D search. Our approach is designed for slowly moving targets, and the accuracy of the time-delay estimate is similar to that of the full batch-wise 2-D method. Given the time-delay, the coherency between batches can be restored when estimating the Doppler parameter. Thereby, the separable approach is found to yield superior Doppler estimates
over a wide parameter range.

In addition to reducing computational complexity, the proposed separable estimation technique also significantly reduces the communication overhead in a distributed radar setting.
\end{abstract}
\section{Introduction}


A network of passive radar sensors takes advantage of existing electromagnetic signals to detect and locate targets. The so-called Illuminators of Opportunity (IO) can be TV or radio transmitters, or even satellites
\cite{kuschel_tutorial_2019}. Such systems have received much attention in the signal processing and radar system community, due to advantages in terms of power consumption and covertness among others \cite{kuschel_tutorial_2019,greco_passive_2023}.

In passive radar, the lack of knowledge of the transmitted signal poses a challenge, see e.g. \cite{tong_cramerrao_2019,zhang_sparsity-based_2019,zhang_maximum_2020} for approaches to jointly estimate this signal and the target parameters. We assume here that the Receiver Nodes (RN) are equipped with a separate Reference Channel (RC), directed towards the IO transmitter. In essence, the (noisy) RC signal then replaces the IO waveform when estimating the target parameters from the Surveillance Channel (SC) data.
Apart from noise in the RC, this scenario resembles an active radar using a random waveform 
\cite{AnkeletalNoiseRadarIET2023} or an integrated sensing and communication (ISAC) application \cite{Zhang_etal_ISAC_2025}.

The case where an RC is available
is considered, e.g. in \cite{gogineni_passive_2018}, where also a technique to handle multipath (clutter) in the RC is proposed, using the principal component of the cross-correlation matrix between the RC and the RC signals. In \cite{zhang_maximum_2019}, the Maximum Likelihood (ML) estimator is derived for this scenario, assuming both channels to be corrupted by White Gaussian Noise (WGN). 
In the present contribution, we consider the case of a single antenna and additive noise in both channels, as well as Direct-Path Interference (DPI) and Clutter Interference (CI) in the SC. Any physical separation between the two channels is ignored, since this can easily be compensated for.
The Extensive Cancellation Approach (ECA) of \cite{colone_multistage_2009} applies an interference cancelation followed by a matched-filtering approach to estimate the target parameters in a 2-D geometry. 
The statistical estimation performance of this approach is presented in \cite{Viberg_etal_CAMSAP2025}.
A similar approach is also pursued in \cite{zhou_direct_2024}, where it is compared to the full ML that uses both channels to estimate the IO signal. 

The above mentioned approaches require a 2-D mapping over the delay and Doppler parameters to be done at each RN and transmitted to the Central Node (CN). 
The purpose of the present contribution is to introduce a computationally attractive suboptimal estimator at each RN, that allows estimating the delay and Doppler parameters separately. 
This yields a 1-D mapping at each RN rather than 2-D, and at the CN the target localization can be done separately from the velocity estimation. Not only does this result in significant computational savings, but also the required data transmission from RNs to the CN is vastly reduced. Our approach is applicable to the detection and localization of slow targets that are located within the range span of CI from stationary scatterers. The focus is on the estimation problem and detection is not addressed herein. 
To the best of our knowledge, no separable delay-Doppler estimator has been presented before for this case. We note, though, that a separable estimator in a different radar array context is presented in 
\cite{swindlehurst_maximum_1998}. Although we assume a single target, the proposed technique can easily be adapted to a multi-target scenario, provided the targets are sufficiently far apart.



\section{Problem Description and Data Model}

We consider a passive radar scenario where $K$ RNs collect data emanating from reflections of an unknown IO signal. 
For simplicity, we assume a single IO and a single target. We also focus primarily on the processing required at each RN rather than at the CN.
The target of interest is moving at constant non-zero speed, and the goal is to estimate its position and velocity. The clutter is assumed to emanate from scatterers that remain stationary during the observation interval. The RNs can transmit data to a CN, and exploiting knowledge of the geometry of the scenario, the combined data are used at the CN to determine the target position and velocity. 
Thus, the positions of the IO and all RNs are assumed to be known at the CN. Further,
the data from different RNs are synchronized to time-delay (relative the bandwidth), but not to phase (relative the carrier), see e.g. \cite{m_weiss_synchronisation_2004}.


The IO transmits a signal $\Re\{s(t)e^{j\omega_{c}t}\}$, where $s(t)$ is the baseband complex envelope and $\omega_c$ the carrier frequency. 
Each RN is equipped with a Reference Channel (RC) and a Surveillance Channel (SC). The RC obtains a copy of the transmitted waveform $s(t)$ at high SNR: $x(t) = a\, s(t) + n(t)$. This is used as a reference in the SC signal, which is expressed as
\begin{align}
    y(t) &= \Tilde{b}s(t) + y^c(t)
    + \Tilde{d}\, s(t-\tau) e^{j\omega t} + \Tilde{e}(t)  \label{eq:surv_s}  \\ 
&=    b\, x(t) +  
    y^c(t)
    + d\, x(t-\tau) e^{j\omega t} + e(t)\,.\label{eq:surv}
\end{align}
In (\ref{eq:surv_s}), the first term is the DPI, the second contains clutter from stationary objects (see further below),
the third is the target reflection, and the last term is receiver noise. 
The factor $e^{j\omega t}$ represents the Doppler effect due to the target motion, which is assumed to be with constant velocity during the data collection interval.
In (\ref{eq:surv}),
we use the RC signal $x(t)$ in lieu of the unknown waveform $s(t)$, so the amplitude parameters, as well as the noise term, are defined relative to the RC amplitude $a$.
The noise term in (\ref{eq:surv}) also accounts for the noise in the RC $x(t)$. 
Since the SNR is assumed to be much higher in the RC channel, this contribution is ignored here. The effect of reference channel noise on estimator performance is further studied in \cite{Viberg_etal_CAMSAP2025}.

The data model (\ref{eq:surv}) is identical for all RNs, but with different amplitudes, time-delays $\tau$ and Doppler frequencies $\omega$, as given by the geometry of the scenario.
We focus first on the required processing at a single RN, and
use (\ref{eq:surv}) as the generic model.
Let the data collection time in the SC be $0\leq t < T$, in which $N$ samples are collected at time instances $t_n = n\Delta T$, $\Delta T = T/N$, $n=0,\dots,N-1$.
We assume a slowly moving target (dozens of m/s), and the number of samples is large $(>10^5)$.
For practical reasons, the data set is divided into $M$ non-overlapping batches that are processed independently as detailed below.
Each batch contains $Q=\lfloor N/M \rfloor$ samples. The $m$th batch of SC data then consists of the $Q\times 1$-vector $\ybf_m$, with elements
$\{y(t_n)\}$ for $n=\{ (m-1)Q,\dots,mQ-1 \}$. From (\ref{eq:surv}), the SC batch data vector is modeled by 
 \begin{equation}
     \ybf_{m} = b_{m} \xbf_m + 
    \ybf_m^c +
    d_{m}\, \xbf_m(\tau) \odot \vbf(\omega) + \ebf_{m}\,, \label{eq:vsurv}
\end{equation}
where 
$\xbf_m$ and $\ebf_m$ are defined conformably with $\ybf_m$ and $\xbf_m(\tau)$ is the vector $\xbf_m$ with a time-delay $\tau$, which is here treated as a real-valued parameter.
Further, the target Doppler is modeled by the length-$Q$ DFT vector $\vbf(\omega)$,
\begin{equation}
    \vbf(\omega) = 
    [1,e^{j\omega \Delta T},  \dots,e^{j\omega (Q-1) \Delta T}]^T,
\end{equation}
and $\odot$ represents the Schur/Hadamard product (elementwise multiplication). 

For a finite-bandwidth signal, the CI $\ybf_k^c$ can be well-approximated using a Finite Impulse Response (FIR) model
$$
\ybf_m^c \approx \sum\limits_{l\in \mathcal{L}}
c_l \xbf_m(l\Delta T)
=\sum\limits_{l=1}^L
c_l \xbf_m(l\Delta T)\,,
$$
where the set $\mathcal{L}$, of cardinality $L$ contains the range of time delays $l$ for which we expect clutter returns. For the sake of simplicity, we assume in the second equality that $\mathcal{L}=\{1,\dots,L\}$.
Hence, the clutter vector in (\ref{eq:vsurv}) is written as
\begin{equation}
    \ybf_m^c = \Xbf_m \cbf_m\,,
\label{eq:clutter}
\end{equation}
where $\Xbf_m$ is a $Q\times L$ Toeplitz matrix containing samples of $x(t_n)$ for 
$(m-1)Q-L \leq n \leq mQ-2$,
and where $\cbf_m = [c_1,\dots,c_L]^T$ is the vector of FIR filter coefficients. 
It is assumed that $L<Q-1$ and that $\Xbf_m$ has full column rank for all $m$.
Due to the time-delay in the clutter filter, there will be some data overlap between the $\Xbf_m$ matrices, and the RC data must be available $L$ samples before the SC begins.


\section{Target Parameter Estimation}

The proposed estimation methods have the following steps, similar to \cite{colone_multistage_2009}. First, each RN uses the RC output 
to cancel the DPI and CI in the SC, and subsequently the target's time-delay and Doppler parameters are estimated. These are fed to the CN, which performs the matching to the target position and velocity parameters. 

\subsection{Baseline Approach: Full 2-D Search}

Inserting (\ref{eq:clutter}) into (\ref{eq:vsurv}), we express the SC data at the $m$:th batch as 
\begin{equation}
     \ybf_{m} = \Xbf_I \fbf_I  +
    d_{m}\, \hat{\abf}_m(\tau,\omega)  + \ebf_{m}\,, \label{eq:vsurv2}
\end{equation}
where we have combined the DPI and the clutter components into the single interference term $\Xbf_I\fbf_I = [\xbf_m,\Xbf_m]\,[b_m^{},\cbf_m^T]^T$. Further, 
the "steering vector" $\hat{\abf}(\tau,\omega) $ is defined by
\begin{equation}
\hat{\abf}_m(\tau,\omega) = \xbf_m(\tau) \odot \vbf(\omega)\,.
\end{equation}
Following \cite{colone_multistage_2009}, see also \cite{Viberg_etal_CAMSAP2025}, a Least-Squares (LS) fit of the 
interference model results in an effective cancellation. Define the orthogonal projection matrix onto the orthogonal complement of the span of 
$\Xbf_I$ as 
\begin{equation}
    \hat{\Pibf}^{\perp} =  \Ibf - \hat{\Pibf}
    =  \Ibf -
    \Xbf_I\left(\Xbf_I^{H}\Xbf_I\right)^{-1}\Xbf_I^H   .
    \label{eq:Piperphat}
\end{equation}
The linear LS fit w.r.t.~$\fbf_m$ then results in the interference-cleaned LS criterion
\begin{equation}
    \ell(d_m,\tau,\omega) =
     \left\| \hat{\Pibf}^{\perp} \left( \ybf_m -
    d_{m}\, \hat{\abf}_m(\tau,\omega) \right) \right\|^2 .
\label{eq:ML3}
\end{equation}
Substituting the minimizing $d_m$ from (\ref{eq:ML3}) back into the criterion then leads to 
a maximization of the following interference-canceled and normalized version of the 2-D delay-Doppler ambiguity function:
\begin{equation}
P_m(\tau,\omega) =
\frac{ |\hat{\abf}_m^H(\tau,\omega) \hat{\Pibf}^{\perp} \ybf_m|^2}{\hat{\abf}_m^H(\tau,\omega)\hat{\Pibf}^{\perp} \hat{\abf}_m(\tau,\omega)} =
\ybf_m^H \hat{\Pbf}_m(\tau,\omega)
\, \ybf_m^{}
\, ,
\label{eq:ML4}
\end{equation}
where 
$\hat{\Pbf}_m(\tau,\omega)$
is the orthogonal projection matrix onto the range space of $\hat{\Pibf}^{\perp}\hat{\bf a}_m(\tau,\omega)\,$:
\begin{equation}
\hat{\Pbf}_m(\tau,\omega)= 
\frac{\hat{\Pibf}^{\perp}\hat{\abf}_m(\tau,\omega) \hat{\abf}_m^H(\tau,\omega)\hat{\Pibf}^{\perp}}{\hat{\abf}_m^H(\tau,\omega)\hat{\Pibf}^{\perp} \hat{\abf}_m(\tau,\omega)} = \Ibf - 
\hat{\Pbf}^{\perp}_m(\tau,\omega)\, .
\label{eq:Pdef}
\end{equation}

In principle, the amplitude parameters are the same in all batches for the $k$th RN. However, the interference cancellation is done in each batch independently to reduce the complexity and allow parallel processing. 
In the baseline approach, the resulting ``power functions" (\ref{eq:ML4}) are then added incoherently to form the aggregated inference statistic
\begin{equation}
\mathcal{P}_k(\tau_k,\omega_k) =
\sum\limits_{m=1}^M P_m(\tau_k,\omega_k)\,,
\label{eq:ML5}
\end{equation}
where we have applied the index $k$ for the $k$th RN.
Note that (\ref{eq:ML5}) is different from the ECA-B approach of \cite{colone_multistage_2009}, where the batches are combined coherently.
Although (\ref{eq:ML5}) depends implicitly on the target position and velocity, the $k$th receiver node can only evaluate the criterion with respect to the target delay and Doppler parameters relative its own position. Thus, at node $k$, (\ref{eq:ML5}) is computed on a $(\tau_k,\omega_k)$-grid, and the ``significant" values are transmitted to the CN for further processing. 

\subsection{Target Localization}

The final step is to combine the delay-Doppler information from all receiver nodes 
at the central node. 
The available data are the sampled versions of (\ref{eq:ML5}) from all nodes. Since the data are independent, they are simply added to form
\begin{equation}
V_{K}(\thetabf) = \sum_{k=1}^K  \mathcal{P}_k(\tau_k(\thetabf),\omega_k(\thetabf))\, ,
\label{eq:GlobalML}
\end{equation}
where $\tau_k=\tau_k(\thetabf)$ and $\omega_k=\omega_k(\thetabf)$ are known functions of the 4-D target parameter vector $\thetabf$, which contains the $(x,y)$ coordinates as well as its speed in the $x$ and $y$ directions. 
The global LS estimator is now to perform a search of (\ref{eq:GlobalML}) over the target parameters in 4 dimensions. 
For each hypothesized target localization and velocity, the corresponding time-delay and Doppler parameters are calculated for each node. The resulting value of (\ref{eq:ML5}) is added to the global ``power function", and if this sample is missing at a particular node, we simply add zero. It should be noted that the discretizations of $(\tau_k,\omega_k)$ at the different RNs are not synchronized, and caution is required when combining their information \cite{yang_multitarget_2022}.
It is also noted that once the target parameters are estimated, a target decision should follow, as in e.g. \cite{yang_multitarget_2022}. 

\subsection{Separable Delay and Doppler Estimation}
\label{subsec:sep_del_doppl}
Recall the data model (\ref{eq:vsurv2}). The terms in the DFT vector $\vbf(\omega)$ are of the form
$e^{jq\omega\Delta T}$ for $q=0,\dots,Q-1$ with $Q=\lfloor N/M \rfloor$. If $|\omega\Delta T|\ll 1/Q$, we can perform a first-order approximation $e^{jq\omega\Delta T}\approx 1+jq\omega\Delta T$. Thus, the steering vector is approximated as
$$
\hat{\abf}_{m}(\tau,\omega) = \xbf_{m}(\tau)\odot \vbf(\omega) \approx
\xbf_{m}(\tau) + \Dbf\, \xbf_{m}(\tau)\, j\omega\Delta T\, ,
$$
where 
$\Dbf = \text{Diag}(0,1,\dots,Q-1)$. 
Next, it is assumed that the target position is within the range of the clutter extent, so that the first term above is annihilated by the interference projection.
Therefore, 
$$
\hat{\Pibf}^\perp\hat{\abf}_{m}(\tau,\omega) \approx \hat{\Pibf}^\perp \Dbf\, \xbf_{m}(\tau)\, j\omega\Delta T\, ,
$$
and the interference-cleaned 2-D delay-Doppler ambiguity function 
(\ref{eq:ML4}) becomes
\begin{align}
P_{m}(\tau,\omega) &= \nonumber
\frac{ |\hat{\abf}_{m}^H(\tau,\omega) \hat{\Pibf}^{\perp} \ybf_{m}|^2}{\hat{\abf}_{m}^H(\tau,\omega)\hat{\Pibf}^{\perp} \hat{\abf}_{m}(\tau,\omega)} \\ &\approx
\frac{ |\xbf_{m}^H(\tau)\Dbf^H \hat{\Pibf}^{\perp} \ybf_{m}|^2}{\xbf_{m}^H(\tau)\Dbf^H  \hat{\Pibf}^{\perp} \Dbf \xbf_{m}(\tau)} 
\, ,
\end{align}
i.e.~a function of $\tau$ only
since the $\omega^2$ factor cancels. This means that we can perform a 1-D search over $\tau$, instead of a 2-D search over $(\tau,\omega)$, as in (\ref{eq:ML5}). 
Further, collecting data $\mathcal{P}_k(\tau_k)$ from all K RNs at the CN, we can determine the target position that maximizes the sum over $k$, similar to (\ref{eq:GlobalML}). But since just $\tau_k$ is involved, the sum is a function of target position only, not velocity. Hence, the target localization is separated from the velocity estimation, which reduces the search from 4-D to 2-D, assuming a planar scenario.

Once the time-delay estimate $\hat{\tau}_k$ is obtained, either locally at node $k$ or globally at the CN, the Doppler parameter $\omega_k$ can be estimated by fixing $\tau_k=\hat{\tau}_k$ in (\ref{eq:ML5}) and searching over $\omega_k$ only. 
However, this can result in a high penalty in terms of statistical performance, since the baseline $Q$ for estimating the Doppler may be significantly reduced compared to the full data length $N$. Therefore, we propose to instead use the estimated target amplitudes in each batch, which are related by $d_m = e^{j\Omega\Delta T m}d_0$, where $d_0$ is the amplitude when $M=1$ and $\Omega = Q\omega$ is the Doppler phase shift from one batch to the next. Provided $|\omega\Delta T|<2\pi/Q$, we can uniquely recover $\omega$ from an estimate of $\Omega$.
From (\ref{eq:ML3}), the estimated target amplitude, for a fixed $\hat{\tau}_k$, is obtained as
\begin{align}
\hat{d}_m &= \frac{
\hat{\abf}_m^H(\omega,\hat{\tau}_k) \hat{\Pibf}^\perp \ybf_m }
{
\hat{\abf}_m^H(\omega,\hat{\tau}_k) \hat{\Pibf}^\perp \hat{\abf}_m(\omega,\hat{\tau}_k)
} \nonumber \\
&\approx \frac{1}{j\omega\Delta T} \,
\frac{
\xbf_m^H(\hat{\tau}_k) \Dbf \hat{\Pibf}^\perp \ybf_m }
{
\xbf_m^H(\hat{\tau}_k) \Dbf  \hat{\Pibf}^\perp \Dbf \xbf_m(\hat{\tau}_k)
} 
\end{align}
Since $\omega$ is unknown, we cannot calculate $\hat{d}_m$. But note that 
$$
\angle \hat{d}_m = \angle \Tilde{d}_m  - \pi/2\,,
$$
where $\Tilde{d}_m$ is the amplitude estimate without normalization
$$
\Tilde{d}_m = 
\xbf_m^H(\hat{\tau}_k) \Dbf \hat{\Pibf}^\perp \ybf_m 
\,.
$$
Hence, the proposed method is to apply Tretter's method \cite{tretter_1985} to $\Tilde{d}_m$, which is to fit a straight line to the (unwrapped) phase sequence $\hat{\phi}_m = \angle \Tilde{d}_m$.
If $\hat{d}_m\approx d_m$, then we have
$$
\hat{\phi}_m\approx \Omega\Delta T m + \angle d_0 + \pi/2\,,
\ m=1,\dots,M.
$$
Therefore, the slope of the linear regression gives $\hat{\Omega}_k$, and then we take $\hat{\omega}_k=\hat{\Omega}_k/Q$.
Since this is an explicit solution for $\hat{\omega}$, the separable approach reduces the computational cost by a factor equal to the number of grid-points in $\omega$ required for the 2-D search.

\section{Numerical Examples}
\begin{figure}[h]

  \centering
  \begin{subfigure}[b]{0.47\textwidth}
    \centering
    \includegraphics[width=\linewidth]{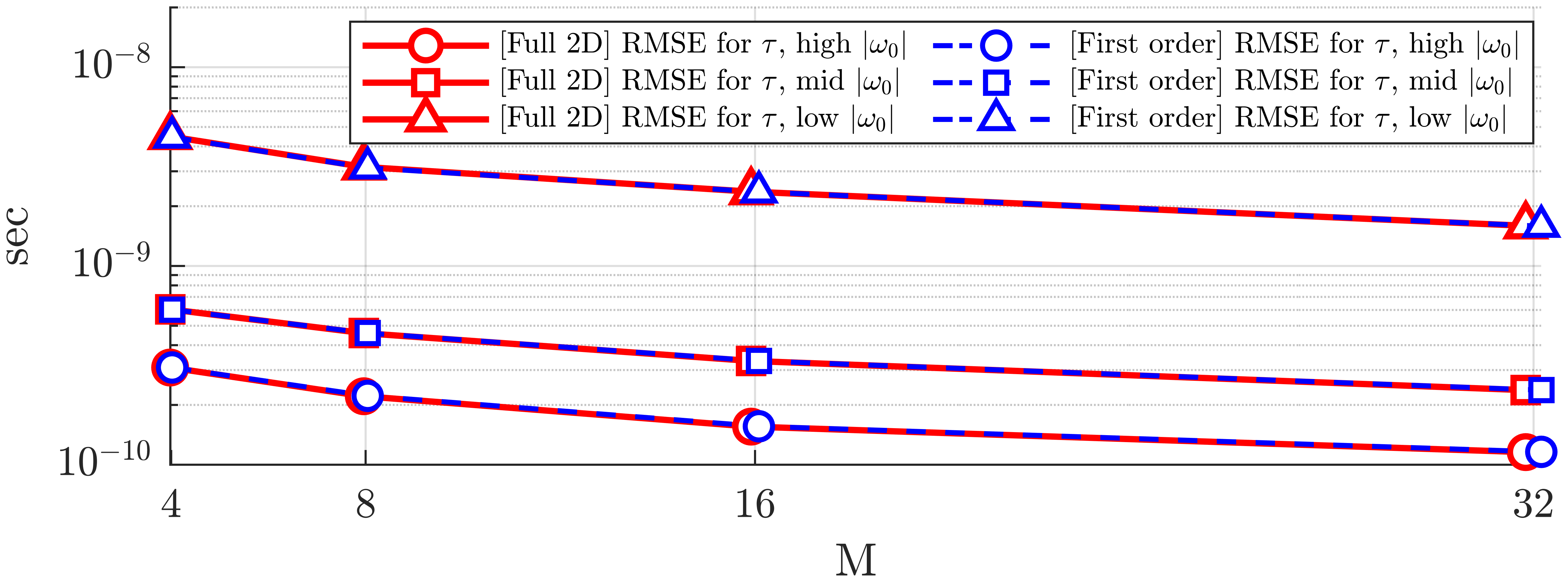}
    \vspace{-15pt}
    \caption{}
     \label{fig:tau_Msweep}
  \end{subfigure}
  
  \begin{subfigure}[b]{0.47\textwidth}
    \centering
    \includegraphics[width=\linewidth]{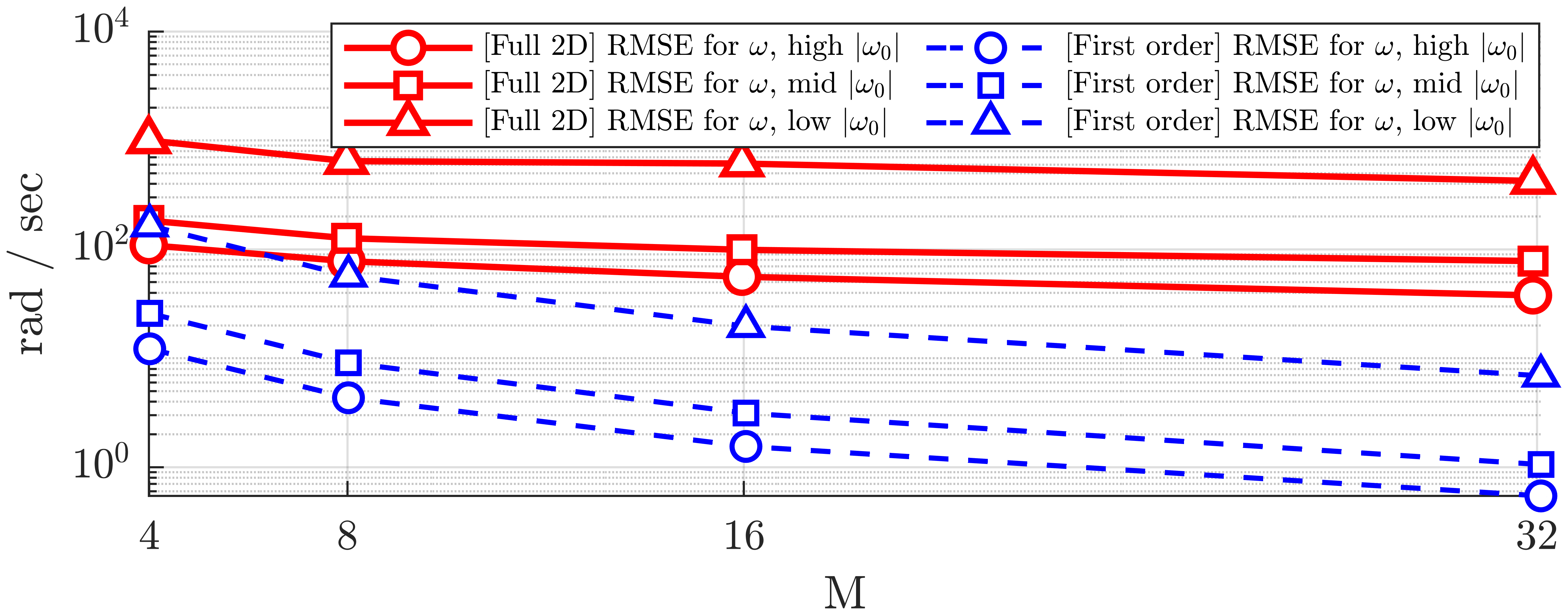}
    \vspace{-15pt}
    \caption{}
    \label{fig:omega_Msweep}
  \end{subfigure}

  \caption{RMS error for a) time-delay and b) Doppler frequency estimates versus the number of batches. The size of each batch is fixed, so $N$ increases with $M$.}
  \label{fig:both_Msweep}
\end{figure}

In this section, we present some numerical performance studies of the proposed method in the form of Monte Carlo simulations. Only a single ``bistatic pair'' (one IO and one RN) is considered, as well as a single target at random location (then kept fixed) within the clutter range. Figure \ref{fig:both_Msweep} displays the impact of the number of data batches \(M\) on the Root Mean Square Error (RMSE) of the estimates. The length of the batches is kept constant, equal to the maximum amount of samples for which a regular laptop can handle the 
interference cancellation, namely \( Q = 2^{14} \) samples (provided that \(L \ll Q\), which in our case was chosen to be \(L = 70 \)). Our focus is on slow targets, and the results for three different $|\omega_0|$ are shown: ``low'' $|\omega_0| \approx 50$ rad/sec, ``mid'' $|\omega_0| \approx 250$ rad/sec and ``high'' $|\omega_0| \approx 450$ rad/sec, where $\omega_0$ is the true Doppler frequency. The sampling rate is \( \Delta T = 4 \times 10^{-8} \) seconds.

The full 2-D search was carried out using the Nelder-Mead algorithm with the ground truth \( (\tau_0, \omega_0) \) \emph{as starting point}, while for the proposed method we started with a coarse grid search along the \( \tau \) dimension only, followed by a finer search within the optimal interval; the estimation of \( \tau_0  \) was then used to retrieve \( \omega_0 \), using the Tretter's regression described in Section \ref{subsec:sep_del_doppl}. We can see that if the batch size is large enough, the 1-D time delay estimations are as good as those in the 2-D incoherent full search, and the latter is outperformed in the Doppler estimations across all the regimes tested.


\begin{figure}[!htbp]

  \centering
  \begin{subfigure}[b]{0.47\textwidth}
    \centering
    \includegraphics[width=\linewidth]{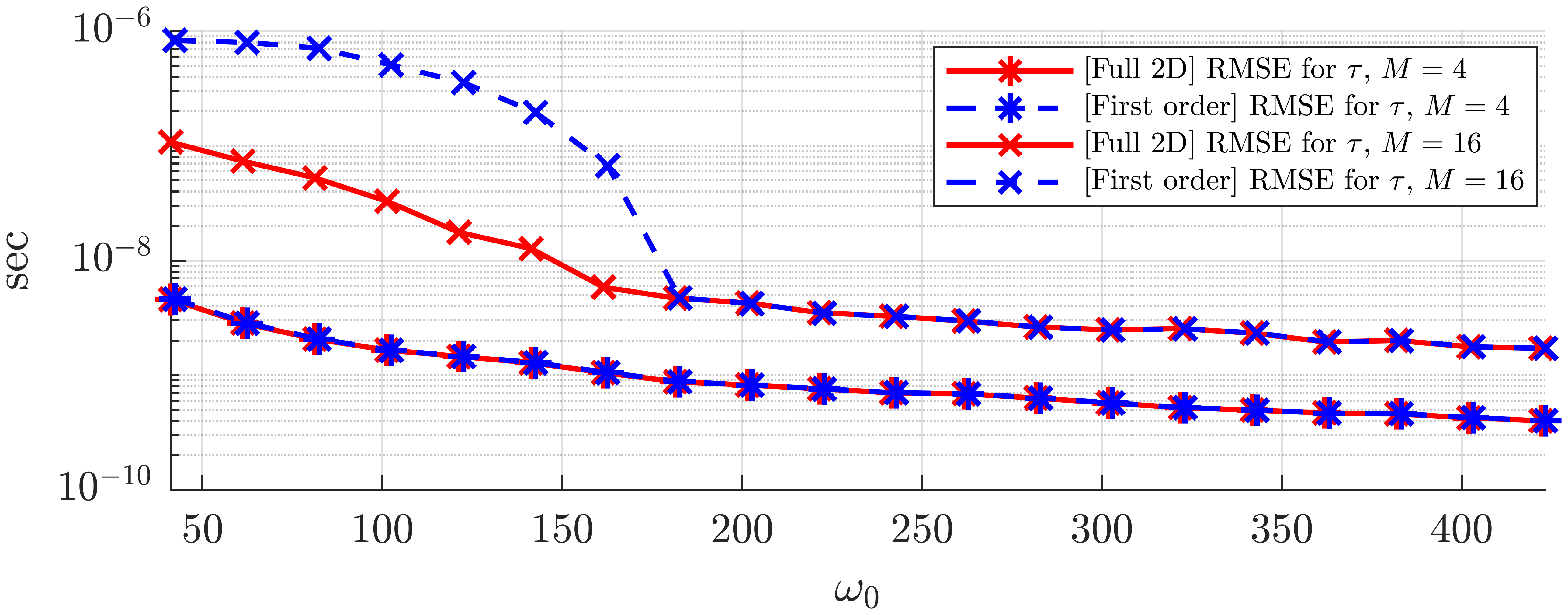}
    \vspace{-15pt}
    \caption{}
     \label{fig:omega_omega0sweep}
  \end{subfigure}
  
  \begin{subfigure}[b]{0.47\textwidth}
    \centering
    \includegraphics[width=\linewidth]{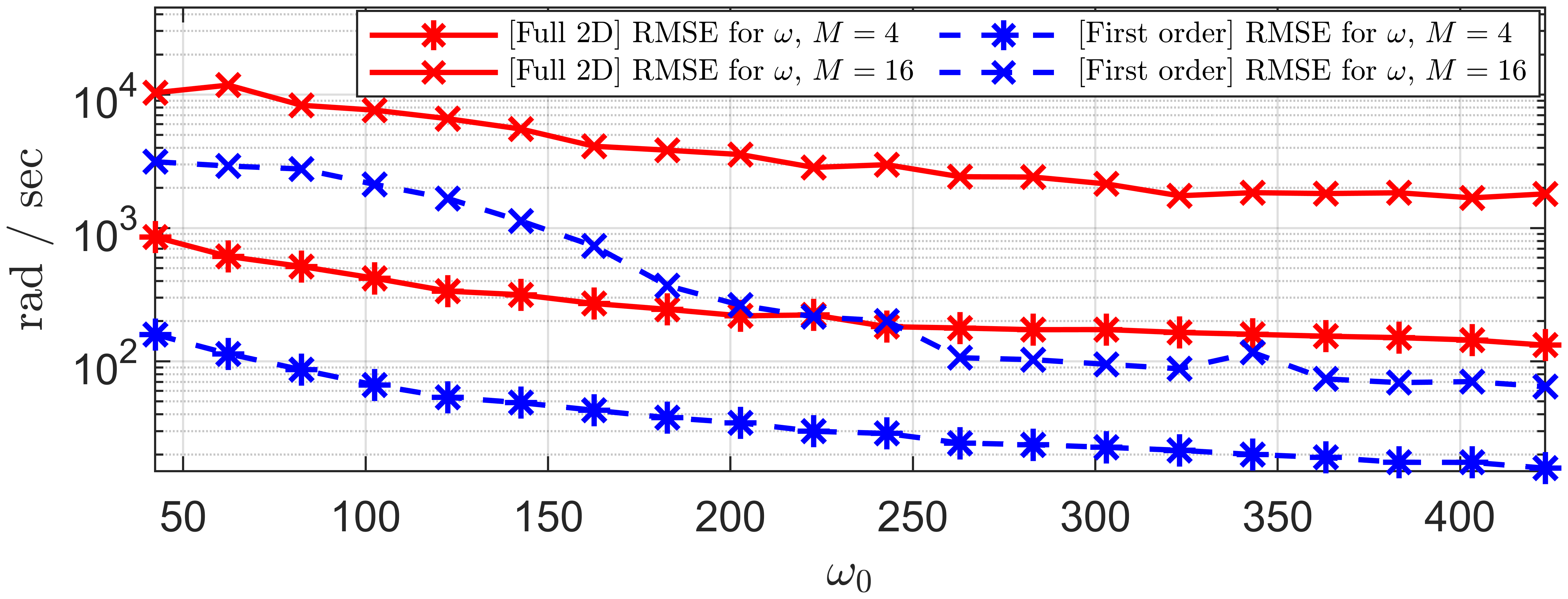}
    \vspace{-15pt}
    \caption{}
    \label{fig:omega_omega0sweep}
  \end{subfigure}
  \caption{RMS error for a) time-delay and b) Doppler frequency estimates versus the target's Doppler frequency. The total data length $N$ is fixed.}
  \label{fig:both_omega0sweep}
\end{figure}

The qualitative notions of ``low'' and ``high'' \( \omega_0 \) are further examined in Figure~\ref{fig:both_omega0sweep}, with the total data length fixed at \( N = 2^{16} \) samples. As expected, there is a threshold effect for slow targets when the batch size is too small (increasing $M$), which is not present for the 2-D approach due to the perfect initialization. 
For large enough batch size,
the performance remains solid also for very small $\omega_0$, and the proposed method outperforms the reference in Doppler estimation.

\section{Conclusions}
In this work we propose a method for time-delay and Doppler-frequency estimation of slowly moving targets using an arbitrary-waveform sensor. Our approach leverages a first-order approximation of the data model to estimate the two parameters separately, yielding significant computational savings, particularly in a distributed radar settings with fast sampling and large data streams. The numerical study indicates that, given a sufficiently large batch size \( Q \), the proposed approach estimates the time-delay parameter as accurately as does the incoherent full 2-D estimator over a wide range of target speeds, while the Doppler estimates are improved for the separable approach across all tested regimes.


%

\bibliographystyle{IEEEbib}
\bibliography{DistributedRadarRefs}

@article{zhang_sparsity-based_2019,
	title = {A {Sparsity}-{Based} {Passive} {Multistatic} {Detector}},
	volume = {55},
	number = {6},
	journal = {IEEE Trans. Aerosp. Electron. Syst.},
	author = {Zhang, Xin and Sward, Johan and Li, Hongbin and Jakobsson, Andreas and Himed, Braham},
	year = {2019},
	pages = {3658--3666},
}

@article{kuschel_tutorial_2019,
	title = {Passive radar tutorial},
	volume = {34},
    number = {2},
	doi = {10.1109/MAES.2018.160146},
	shorttitle = {Tutorial},
	pages = {2--19},
	journal = {{IEEE} Aerosp. Electron. Syst. Mag.},
	author = {Kuschel, Heiner and Cristallini, Diego and Olsen, Karl Erik},
	urldate = {2025-06-17},
	year = {2019},
	langid = {english},
}

@article{tong_cramerrao_2019,
	title = {Cramér–Rao Lower Bound Analysis for Stochastic Model Based Target Parameter Estimation in Multistatic Passive Radar With Direct-Path Interference},
	volume = {7},
	rights = {https://creativecommons.org/licenses/by/4.0/legalcode},
	issn = {2169-3536},
	url = {https://ieeexplore.ieee.org/document/8753487/},
	doi = {10.1109/ACCESS.2019.2926353},
	pages = {106761--106772},
	journaltitle = {{IEEE} Access},
	journal = {{IEEE} Access},
	author = {Tong, Jing and Gaoming, Huang and Wei, Tian and Huafu, Peng},
	urldate = {2025-06-18},
	year = {2019},
	langid = {english},
}

@article{zhang_maximum_2020,
	title = {Maximum Likelihood and {IRLS} Based Moving Source Localization with Distributed Sensors},
	rights = {https://ieeexplore.ieee.org/Xplorehelp/downloads/license-information/{IEEE}.html},
	issn = {0018-9251, 1557-9603, 2371-9877},
	url = {https://ieeexplore.ieee.org/document/9188024/},
	doi = {10.1109/TAES.2020.3021809},
	pages = {448--461},
    volume = {57},
    number = {1},
	journaltitle = {{IEEE} Transactions on Aerospace and Electronic Systems},
	journal = {{IEEE} Trans. Aerosp. Electron. Syst.},
	author = {Zhang, Xudong and Wang, Fangzhou and Li, Hongbin and Himed, Braham},
	urldate = {2025-06-18},
	year = {2020},
	langid = {english},
}

@article{gogineni_passive_2018,
	title = {Passive Radar Detection With Noisy Reference Channel Using Principal Subspace Similarity},
	volume = {54},
	rights = {https://ieeexplore.ieee.org/Xplorehelp/downloads/license-information/{IEEE}.html},
	issn = {0018-9251, 1557-9603, 2371-9877},
	url = {https://ieeexplore.ieee.org/document/7990183/},
	doi = {10.1109/TAES.2017.2730998},
	pages = {18--36},
	number = {1},
	journaltitle = {{IEEE} Transactions on Aerospace and Electronic Systems},
	shortjournal = {{IEEE} Trans. Aerosp. Electron. Syst.},
	author = {Gogineni, Sandeep and Setlur, Pawan and Rangaswamy, Muralidhar and Nadakuditi, Raj Rao},
	urldate = {2025-06-16},
	year = {2018},
}

@article{yang_multitarget_2022,
	title = {Multitarget {Detection} {Strategy} for {Distributed} {MIMO} {Radar} {With} {Widely} {Separated} {Antennas}},
	volume = {60},
	journal = {IEEE Trans. Geosci. Remote Sensing},
	author = {Yang, Shixing and Yi, Wei and Jakobsson, Andreas},
	year = {2022},
	pages = {1--16},
}

@inproceedings{m_weiss_synchronisation_2004,
	address = {Anchorage, AK, USA},
	title = {Synchronisation of bistatic radar systems},
	volume = {3},
	isbn = {978-0-7803-8742-3},
	url = {http://ieeexplore.ieee.org/document/1370671/},
	doi = {10.1109/IGARSS.2004.1370671},
	language = {en},
	urldate = {2025-04-28},
	booktitle = {{IEEE} {International} {IEEE} {International} {IEEE} {International} {Geoscience} and {Remote} {Sensing} {Symposium}, 2004. {IGARSS} '04. {Proceedings}. 2004},
	publisher = {IEEE},
	author = {{M. Weiss}},
	year = {2004},
	pages = {1750--1753},
}

@incollection{greco_passive_2023,
	address = {Cham},
	title = {Passive {Radar}: {A} {Challenge} {Where} {Resourcefulness} {Is} the {Key} to {Success}},
	isbn = {978-3-031-21974-0 978-3-031-21975-7},
	shorttitle = {Passive {Radar}},
	url = {https://link.springer.com/10.1007/978-3-031-21975-7_8},
	language = {en},
	urldate = {2025-04-28},
	booktitle = {Women in {Telecommunications}},
	publisher = {Springer International Publishing},
	author = {Filippini, Francesca and Colone, Fabiola},
	editor = {Greco, Maria Sabrina and Cassioli, Dajana and Ullo, Silvia Liberata and Lyons, Margaret J.},
	year = {2023},
	doi = {10.1007/978-3-031-21975-7_8},
	note = {Series Title: Women in Engineering and Science},
	pages = {223--247},
}

@article{colone_multistage_2009,
	title = {A {Multistage} {Processing} {Algorithm} for {Disturbance} {Removal} and {Target} {Detection} in {Passive} {Bistatic} {Radar}},
	volume = {45},
	number = {2},
	journal = {IEEE Trans. Aerosp. Electron. Syst.},
	author = {Colone, F. and O'Hagan, D. W. and Lombardo, P. and Baker, C. J.},
	year = {2009},
	pages = {698--722},
}

@article{zhou_direct_2024,
	title = {Direct {Target} {Localization} for {Distributed} {Passive} {Radars} {With} {Direct}-{Path} {Interference} {Suppression}},
	volume = {72},
	journal = {IEEE Trans. Signal Process.},
	author = {Zhou, Qiyu and Yuan, Ye and Venturino, Luca and Yi, Wei},
	year = {2024},
	pages = {3611--3625},
}

@article{zhang_maximum_2019,
	title = {Maximum {Likelihood} {Delay} and {Doppler} {Estimation} for {Passive} {Sensing}},
	volume = {19},
	copyright = {https://ieeexplore.ieee.org/Xplorehelp/downloads/license-information/IEEE.html},
	issn = {1530-437X, 1558-1748, 2379-9153},
	url = {https://ieeexplore.ieee.org/document/8489883/},
	doi = {10.1109/JSEN.2018.2875664},
	language = {en},
	number = {1},
	urldate = {2025-04-28},
	journal = {IEEE Sensors J.},
	author = {Zhang, Xudong and Li, Hongbin and Himed, Braham},
	month = jan,
	year = {2019},
	pages = {180--188},
}

@article{swindlehurst_maximum_1998,
	title = {Maximum likelihood methods in radar array signal processing},
	volume = {86},
	copyright = {https://ieeexplore.ieee.org/Xplorehelp/downloads/license-information/IEEE.html},
	issn = {00189219},
	url = {http://ieeexplore.ieee.org/document/659495/},
	doi = {10.1109/5.659495},
	language = {en},
	number = {2},
	urldate = {2025-08-29},
	journal = {Proceedings of the IEEE},
	author = {Swindlehurst, A.L. and Stoica, P.},
	month = feb,
	year = {1998},
	pages = {421--441},
}

@inproceedings{Viberg_etal_CAMSAP2025,
	address = {Dominican Republic},
	title = {STATISTICAL ANALYSIS OF TARGET PARAMETER ESTIMATION USING PASSIVE RADAR},
    publisher = {IEEE},
    booktitle = {Proc. 10th Int'l Conf. Computational Advances in Multi-Sensor Adaptive Processing},
	author = {M. Viberg and D. Gerosa and T. McKelvey and T. Eriksson},
	year = {2025},
	note = {To Appear}
}

@article{tretter_1985,
	title = {Estimating the frequency of a noisy sinusoid by linear regression (Corresp.)},
	volume = {31},
	issn = {00189219},
	url = {https://ieeexplore.ieee.org/abstract/document/1057115},
	language = {en},
	number = {6},
	journal = {IEEE Transactions on Information Theory},
	author = {Tretter, S.},
	month = nov,
	year = {1985},
	pages = {832--835}
}

@ARTICLE{Zhang_etal_ISAC_2025,
  author={Zhang, Zhenkun and Ren, Hong and Pan, Cunhua and Hong, Sheng and Wang, Dongming and Wang, Jiangzhou and You, Xiaohu},
  journal={IEEE Transactions on Communications}, 
  title={Target Localization in Cooperative ISAC Systems: A Scheme Based on 5G NR OFDM Signals}, 
  year={2025},
  volume={73},
  number={5},
  pages={3562-3578},
  doi={10.1109/TCOMM.2024.3486981}}

@ARTICLE{AnkeletalNoiseRadarIET2023,
  author={Martin Ankel and Robert Jonsson and Tomas Bryllert and Lars M. H. Ulander and Per Delsing},
  journal={IET Radar, Sonar \& Navigation}, 
  title={Bistatic noise radar: Demonstration of correlation noise suppression}, 
  year={2023},
  volume={17},
  number={3},
  pages={351-361},
}

\end{document}